\documentclass{article}
\usepackage{ijcai23}

\usepackage{times}
\usepackage{soul}
\usepackage{url}
\usepackage[hidelinks]{hyperref}
\usepackage[utf8]{inputenc}
\usepackage[small]{caption}
\usepackage{graphicx}
\usepackage{amsmath}
\usepackage{amsthm}
\usepackage{booktabs}
\usepackage{algorithm}
\usepackage{algorithmic}
\usepackage[switch]{lineno}
\usepackage{color}
\usepackage{xcolor}
\usepackage{multirow}
\usepackage{rotating}
\usepackage{array}
\usepackage{booktabs}
\usepackage{enumitem}
\newcommand{\citet}[1]{\citeauthor{#1}~\shortcite{#1}}

\title{A Survey on User Behavior Modeling in Recommender Systems}

\author{
Zhicheng He$^{1*}$\and
Weiwen Liu$^{1*}$\and
Wei Guo$^{1}$\and
Jiarui Qin$^2$\and
Yingxue Zhang$^3$\and \\
Yaochen Hu$^3$\and
Ruiming Tang$^{1\dagger}$
\affiliations
$^1$Huawei Noah’s Ark Lab, Shenzhen, China\\
$^2$Shanghai Jiao Tong University, Shanghai, China\\
$^3$Huawei Noah’s Ark Lab, Montreal, Canada
\emails
\{hezhicheng9, liuweiwen8, guowei67, yingxue.zhang, yaochen.hu, tangruiming\}@huawei.com,
qinjr96@sjtu.edu.com
}
\begin{document}

\maketitle

\begin{abstract}
User Behavior Modeling (UBM) plays a critical role in user interest learning, which has been extensively used in recommender systems.
Crucial interactive patterns between users and items have been exploited, which brings compelling improvements in many recommendation tasks.
In this paper, we attempt to provide a thorough survey of this research topic.
We start by reviewing the research background of UBM.
Then, we provide a systematic taxonomy of existing UBM research works, which can be categorized into four different directions including \textit{Conventional UBM, Long-Sequence UBM, Multi-Type UBM, and UBM with Side Information.}
Within each direction, representative models and their strengths and weaknesses are comprehensively discussed.
Besides, we elaborate on the industrial practices of UBM methods with the hope of providing insights into the application value of existing UBM solutions.
Finally, we summarize the survey and discuss the future prospects of this field.
\let\thefootnote\relax\footnotetext{* Zhicheng He and Weiwen Liu are the co-first authors, and Ruiming Tang is the corresponding author.}
\end{abstract}

\section{Background}\label{sec:background}
With the rapid development of various Internet applications, Recommender Systems (RS) have become increasingly indispensable to both personalized services and the alleviation of information overloading \cite{Zhang2021Deep}.
One major bottleneck of RS research is the scarcity of explicit feedback regarding users' preferences \cite{Hu2008Collaborative}.
Instead, user preferences are implicitly recorded in the coarse and noisy behavior logs.
To ameliorate this bottleneck, many researchers have been devoted to the study of User Behavior Modeling (UBM), aiming to explore and exploit user interest representations from the behavior history \cite{Zhang2021Deep}.

\begin{figure}[t!]
    \centering
    \includegraphics[width=\columnwidth]{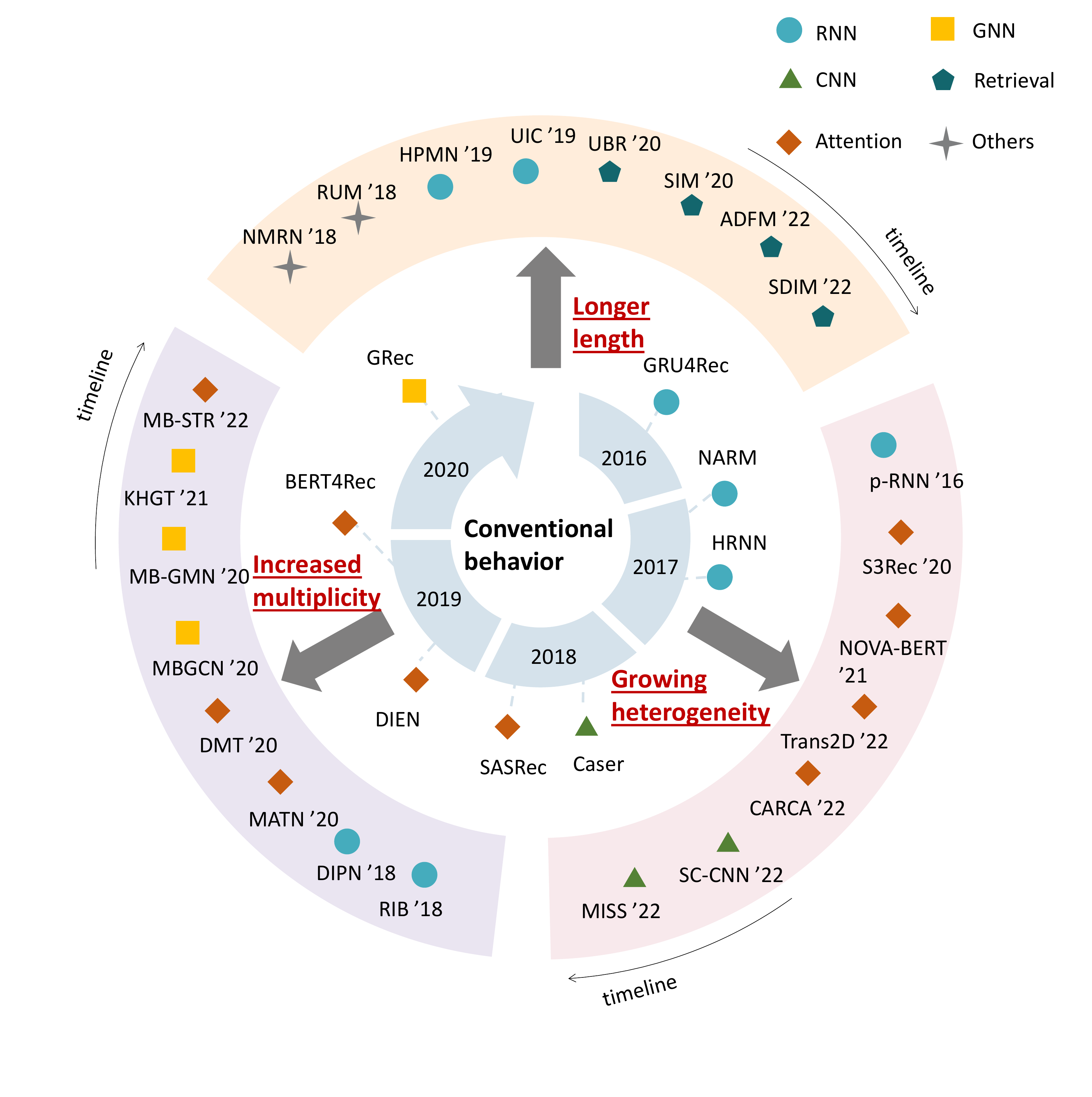}
    \vspace{-12mm}
    \caption{Taxonomy for user behavior modeling. The primary technique of each method is marked with a dyed shape.}
    \label{fig:orgchart}
    \vspace{-5mm}
\end{figure}

Despite the different forms of application tasks (e.g., ranking or next-item prediction), the common learning objective of recommendation models is to predict given users' interests in particular items.
Specifically, for a UBM based recommendation model $F_{\Theta}^{UBM}(\cdot)$ parameterized by $\Theta$, the interest $\text{P}(u, i)$ of a given user $u$ for a target item $i$ is predicted as
\begin{equation}\label{eq:form}
\text{P}(u, i) = F_{\Theta}^{UBM}(u, i, \mathcal{H}_u, f_{u}, f_{i}, f_{c}), \ \forall \ u \in \mathcal{U}, \ i \in \mathcal{I}
\end{equation}
where $\mathcal{U}$ and $\mathcal{I}$ are the universal sets of users and items respectively, and $\mathcal{H}_u$ is the user behavior history.
$f_{u}$, $f_{i}$, and $f_{c}$ are the non-behavior user features (e.g. age), item features (e.g. price), and context features (e.g., weekday) respectively.
We do not discuss $f_{u}$, $f_{i}$, and $f_{c}$ as they are not the key concerning issue of UBM, interested readers can refer to \cite{Zhang2021Deep}.
The core research issue of UBM is how to extract and model user interests from the behavior history $\mathcal{H}_u$.
Within $\mathcal{H}_u$, each behavior record $h_{u,k} = \{v_k, t_k, f_{k}\}$ consists of the interacted item $v_k \in \mathcal{I}$, the time stamp $t_k$, and some related features $f_{k}$ (e.g., behavior type and item description).
As presented in Figure \ref{fig:orgchart}, various UBM methods have been proposed to learn effective user interest representations from $\mathcal{H}_u$.
Despite there being already some survey papers on conventional UBM \cite{Wang2019Sequential,Fang2020Deep}, however, they fail to cover the most recent development in user behavior modeling.
In this survey, we will comprehensively analyze the latest UBM research progresses and discusses their advantages and weaknesses from both academic and industrial perspectives.
Furthermore, we lay out three new research trends, i.e., the \textit{longer length},\textit{ increased multiplicity}, and \textit{growing heterogeneity} of user behavior sequences, as illustrated in Figure \ref{fig:orgchart}.

The rest of this paper is organized as follows.
In Section \ref{sec:taxonomy}, we pick out some representative UBM works, categorize the proposed ideas, and analyze their differences and similarities.
On top of that, we propose a new taxonomy of UBM which consists of four research directions as shown in Figure \ref{fig:orgchart}.
The detailed discussion about the achievements, advantages and disadvantages of the four directions are presented in Section \ref{sec:con-ubm}, \ref{sec:long-ubm}, \ref{sec:mb-ubm}, and \ref{sec:side-ubm} respectively.
In Section \ref{sec:industrial}, we further discuss the application value of UBM in large-scale industrial recommender systems.
Finally, Section \ref{sec:future} concludes this survey and sheds some light on the future research directions of UBM.

\section{Taxonomy}\label{sec:taxonomy}

\begin{table*}[t]
\centering
\begin{tabular}{llll}
\hline
\multicolumn{1}{l|}{Branch} & \multicolumn{1}{l|}{Representative} & \multicolumn{1}{l|}{Research Motivation} & Key Techniques\\
\hline
\multicolumn{1}{l|}{\multirow{4}{*}{Conventional}}
& \multicolumn{1}{l|}{GRU4Rec} & \multicolumn{1}{l|}{Learn evolving behavior patterns within session} & RNN \\
\multicolumn{1}{l|}{}
& \multicolumn{1}{l|}{HRNN} & \multicolumn{1}{l|}{Integrate in- \& cross-session behavior patterns} & RNN\\ 
\multicolumn{1}{l|}{}                   
& \multicolumn{1}{l|}{Caser} & \multicolumn{1}{l|}{Learn behavior patterns from different scopes and levels} & CNN \\ 
\multicolumn{1}{l|}{}                   
& \multicolumn{1}{l|}{SASRec} & \multicolumn{1}{l|}{Learn adaptive weights for behavior dependencies} & Attention, Transformer\\ 
\hline

\multicolumn{1}{l|}{\multirow{2}{*}{Long-sequence}} 
& \multicolumn{1}{l|}{UIC} & \multicolumn{1}{l|}{Memorize user interests for long histories ($>1000$)} & Memory Network\\
\multicolumn{1}{l|}{}                   
& \multicolumn{1}{l|}{UBR} & \multicolumn{1}{l|}{Retrieve relevant behaviors from long  histories ($>1000$)} & Retrieval, Ranking \\ 
\hline

\multicolumn{1}{l|}{\multirow{3}{*}{Multi-type}} 
& \multicolumn{1}{l|}{DMT} & \multicolumn{1}{l|}{Efficiently combine user interests from different behaviors} & Transformer, GCN\\
\multicolumn{1}{l|}{}                   
& \multicolumn{1}{l|}{MBGCN} & \multicolumn{1}{l|}{Fuse fine-grained intra- \& cross-type behavior semantics} & Transformer, GCN\\ 
\multicolumn{1}{l|}{}                   
& \multicolumn{1}{l|}{NMTR} & \multicolumn{1}{l|}{Jointly predict different types of user behaviors} & Cascaded prediction \\ 
\hline

\multicolumn{1}{l|}{\multirow{3}{*}{Side information}} 
& \multicolumn{1}{l|}{TiSASRec}  & \multicolumn{1}{l|}{Consider the time interval between two behaviors} & Attention\\
\multicolumn{1}{l|}{}                   
& \multicolumn{1}{l|}{p-RNN}     & \multicolumn{1}{l|}{Consider the multi-modal text and image information} & RNN \\ 
\multicolumn{1}{l|}{}                   
& \multicolumn{1}{l|}{S3Rec}     & \multicolumn{1}{l|}{Extract supervised signals from side information} & SSL, Pre-training \\ 
\hline
\multicolumn{4}{c}{GRU4Rec \cite{gru4rec}, HRNN \cite{hrnn}, Caser \cite{Tang2018Personalized}, UIC \cite{Pi2019Practice}, } \\
\multicolumn{4}{c}{SASRec \cite{Kang2018Self}, UBR \cite{Qin2020User}, DMT \cite{Gu2020Deep}, MBGCN \cite{Jin2020Multi},} \\
\multicolumn{4}{c}{NMTR \cite{Gao2019Neural}, TiSASRec \cite{li2020time}, p-RNN \cite{hidasi2016parallel}, S3Rec \cite{zhou2020s3}.} \\
\hline
\end{tabular}
\vspace{-3mm}
\caption{A taxonomy for existing UBM works according to the research issues they are proposed to solve.}
\label{tab:works}
\vspace{-5mm}
\end{table*}

Various UBM methods have been proposed to advance the exploration of implicit user interest modeling.
To better understand the developing venation of UBM, we identify some representative and influential research works, analyze their research motivations, and summarize their key technical contributions.
As presented in Table \ref{tab:works}, this survey establishes a novel taxonomy for UBM that divides the existing research works into four major categories.
Originating from the conventional UBM research on simple user behaviors, researchers have extended to further consider the \textit{longer length, increased multiplicity, and growing heterogeneity} of user behavior histories, as illustrated in Figure \ref{fig:orgchart}.
We briefly explain the core ideas of the three recent UBM research trends as follows.

\begin{itemize}[leftmargin=*]
    \item \textbf{Conventional UBM} focuses on learning user interest representations from simple historical behavior sequences. 
    The user behavior history $\mathcal{H}_u$ is first processed into a chronologically arranged item sequence with restricted length.
    Subsequently, researchers attempt to explore the complicated behavior patterns from various angles, such as the session structure \cite{gru4rec}, behavior granularity and influence scope \cite{Tang2018Personalized}, and pairwise dependency \cite{Kang2018Self}.
    
    \item \textbf{Long-Sequence UBM} extends conventional UBM by learning from longer sequences of user behavior records (at least in the scale of thousands).
    As more behavior records in $\mathcal{H}_u$ are kept, the model will have the potential to extract richer and long-term user interests.
    However, learning very long sequences with deep models is challenging and have potential computation bottleneck.
    Thus researchers either adopt memory-augmented methods to store long-range behavior dependencies \cite{Pi2019Practice} or take retrieval-based methods to pick out target-relevant behavior histories \cite{Qin2020User}.
    
    \item \textbf{Multi-Type UBM} further extends to study the multiplicity of user behaviors.
    In a unified recommendation model, the explicit modeling of different behavior types (e.g. click and purchase) provides a new perspective to understand behavior patterns and behavior correlations at a fine-grained granularity. 
    Researchers have devoted themselves to a series of new challenges raised by the behavior multiplicity, such as the multi-behavior definition \cite{Guo2019Buying}, multi-behavior fusion \cite{Jin2020Multi}, and multi-behavior prediction \cite{Gao2019Neural}.
    
    \item \textbf{UBM with Side Information} further takes into account the heterogeneous features associated with behavior records.
    The side information helps to recover the contextual interactive logic when behavior records happen, which provides important supplements for understanding complicated user interests.
    Advanced NLP and CV models have been adopted to transform the rich multi-modal information into the user interest representation space \cite{liu2021noninvasive,singer2022sequential}.
\end{itemize}
The four research directions are closely related rather than mutually exclusive, and they enable the fast progression of the UBM research field jointly.
In the following sections, we further elaborate on the four UBM research directions in terms of the key research challenges, representative solutions, and developing trends.

\section{Conventional UBM}\label{sec:con-ubm}
Conventional user behavior modeling focuses on extracting item dependencies and correlations from relatively short-term behavior sequences with the same behavior type \cite{Kang2018Self,Zhou2018Deep}. 
Typical models can be formalized as:
\begin{equation}\label{eq:con-form}
\text{P}(u, i) = F_{\Theta}^{UBM}(u, i, \mathcal{H}_u^{S}), \ \forall \ u \in \mathcal{U}, \ i \in \mathcal{I},
\end{equation}
where $\mathcal{H}_u^{S}$ is the behavior sequence extracted from $\mathcal{H}_u$.
To both ensure the relevancy of extracted behavior records and to reduce model complexity, an upper bound value is used to restrict the length of the user behavior sequence $\mathcal{H}_u^{S}$.
Various deep network structures have been adopted to learn from $\mathcal{H}_u^{S}$, including Recurrent Neural Networks (RNNs), Convolutional Neural Networks (CNNs), and Attention mechanisms.

\subsection{RNN-based Methods}
RNNs naturally capture the long- and short-term dependencies of a sequence. As one of the earliest attempts at user behavior modeling, GRU4Rec \cite{gru4rec} uses Gated Recurrent Units (GRUs) to learn the evolving patterns for user behaviors within a session. Follow-up studies extend GRU4Rec with data augmentation techniques \cite{improved_gru4rec} or improved ranking loss functions \cite{gru4rec+}. NARM \cite{narm} applies an RNN-based encoder-decoder model that combines both global and local interests in a session. \citet{hrnn} further propose a hierarchy of two GRUs to model the evolving of user behaviors within and across sessions, respectively.

\subsection{CNN-based Methods}
\citet{Tang2018Personalized} point out that RNN-based methods may not be able to capture the \textit{skip behaviors} well, where the next step is influenced by the behaviors a few steps earlier rather than the adjacent behavior. Therefore, they propose a Case model that views recent behaviors as an “image” among time and latent dimensions, and learns the sequential patterns and skip behaviors with horizontal and vertical convolutional filters. NextItNet \cite{nextitnet} introduces a generative CNN model with the residual block structure to capture both short- and long-range item dependencies. 

\subsection{Attention-based Methods} 
Recently, attention mechanisms are largely adopted in UBM because of their advantage in modeling interactions between any pair of behaviors, without degradation over the encoding distance \cite{attention}. 
To exploit more supervised information, SASRec \cite{Kang2018Self} adopts self-attention to identify the importance of past behaviors in an auto-regressive prediction framework.
DIN \cite{Zhou2018Deep} applies the attention mechanisms to adaptively learn the representation of user interests from historical behaviors with respect to a certain item. 
However, DIN ignores the evolving character of user interests, which is solved later in DIEN \cite{Zhou2019Deep} with the combination of attention and GRU.
To further consider the intrinsic session structure of user behaviors, DSIN \cite{Feng2019Deep} uses self-attention and Bi-LSTM to extract in-session interest representations and cross-session evolution patterns respectively.
Despite the above achievements, researchers also notice the defects of conventional attention mechanisms in exploring complicated behavior patterns, thus resorting to more sophisticated attention architectures such as the Transformer encoder in BST \cite{Chen2019Behavior} and the stacked bidirectional Transformer encoder in BERT4Rec \cite{Sun2019BERT4Rec}.

\subsection{Discussion}
Research on conventional UBM explores different network structures for extracting dependency patterns from the simple behavior sequence. 
Despite the network structures mentioned above, other deep network structures (e.g., MLP \cite{wang2015learning} and GNN \cite{Chang2021Sequential}) or a combination of multiple network structures \cite{Zhou2019Deep,Feng2019Deep} have also been investigated in conventional UBM. 
For example, SURGE \cite{Chang2021Sequential} constructs item-item interest graphs from behavior sequences using metric learning. 
Different deep techniques are good at capturing different behavior dependencies.
We observe that the captured dependencies in UBM are becoming more complex and practical. 
The dependency modeling progresses from simple unidirectional dependencies (by RNNs) to skip behavior dependencies (by CNNs), and finally to arbitrary pairwise dependency modeling (Attention) or multiple relationships modeling (GNNs).

\section{Long-Sequence UBM}\label{sec:long-ubm}

With the ever-growing volume of data from user-platform interactions, users accumulate a large amount of behavior data \cite{Ren2019Lifelong}. 
Extending from the conventional UBM, long-sequence UBM further attempts to reserve a longer range of user behavior histories, which can be defined as
\begin{equation}\label{eq:long-form}
\text{P}(u, i) = F_{\Theta}^{UBM}(u, i, \mathcal{H}_u^{L}), \ \forall \ u \in \mathcal{U}, \ i \in \mathcal{I},
\end{equation}
where $\mathcal{H}_u^{L}$ reserves as many behavior records in $\mathcal{H}_u$ as possible.
Long-sequence UBM enables recommender systems to further take advantage of long-term behavior dependencies and the periodicity of user behaviors.
However, the user behavior sequences could become very long (at least in thousands), which might be prohibitively expensive to model all the historical user behavior. 
Longer sequences usually incline to contain more noise \cite{Pi2020Search,Qin2020User}, 
and the strict online latency requirements further discourage incorporating long behavior sequences in large-scale industrial production.
To ameliorate the above problems, many research works are proposed from both academia and industry, which can be divided into two major directions: memory-augmented methods and user behavior retrieval-based methods.

\subsection{Memory-Augmented Methods}
Memory-augmented networks are originally proposed in Natural Language Processing (NLP) tasks which explicitly store the extracted knowledge into external memory \cite{Graves2014Neural,Kumar2016Ask}. 
In recommendation scenarios, the external memory stores the user interest representations, which are read and updated by a tailored neural network according to the user behavior records.

NMRN \cite{Wang2018Neural} maintains an external memory bank of several interest representations for each user, which are updated according to the newly generated user behaviors.
The interest representations are associated with personalized weights to reflect users' different tastes.
RUM \cite{Chen2018Sequential} stores the users' historical behaviors explicitly. It uses a first-in-first-out mechanism to maintain the latest interacted items in the user memory.
KSR \cite{Huang2018Improving} takes the item attributes as keys to index the extracted knowledge from the external memory, which enhances the item representations for sequential recommendations.

To handle the long-sequence user behaviors more efficiently and effectively, specifically designed memory-augmented networks are proposed to summarize long behavior sequences into dense vectors.
\citet{Ren2019Lifelong} propose the lifelong sequential modeling framework with the Hierarchical Periodic Memory Network (HPMN).
To capture the multi-scale sequential patterns, a multi-layer GRU architecture is designed to incrementally update external user interest representations with different update frequencies at each layer.
With similar motivations, \citet{Pi2019Practice} design a User Interest Center (UIC) module to store the multi-faceted user interests captured by the Multi-channel user Interest Memory Network (MIMN).
At the inference time, user behavior representations are directly obtained from UIC without online calculation, which decouples the time-consuming modeling procedure from the real-time prediction process.
UIC provides a systematic solution for long-sequence UBM from the industrial perspective and it has successfully deployed in a real-world recommender system.

\subsection{User Behavior Retrieval Methods}
Apart from storing user interests in an external memory, retrieving the most relevant and important behaviors from a very long sequence is an alternative solution. 
The retrieval process could be conducted efficiently and the retrieval step can reduce the potential noise inside the sequence. 
UBR \cite{Qin2020User} uses search engine techniques to retrieve the most relevant behaviors with the target item. 
Instead of using a long consecutive behavior sequence, only a small retrieved set will be used in the following computation. 
\citet{Qin2023Learning} further study how to optimize the retrieval function and propose a learning-to-rank-based optimization method.
SIM \cite{Pi2020Search} proposes hard search and soft search approaches. 
For the hard search, it uses the user ID and item category to build the two-layer index. 
As for the soft search, SIM utilizes local sensitive hashing (LSH) to quickly fetch the relevant behaviors based on their embeddings.
The above two methods are both two-phase models which means the retrieval function is trained separately from the prediction module.
ETA \cite{Chen2021End}, on the contrary, proposes to train the retrieval function in an end-to-end manner.
It uses the SimHash algorithm to map the behavior embeddings and the target item to binary signatures. And based on the binary signatures, the behaviors with the smallest Hamming distance to the target item are retrieved.
Item embeddings are the only parameters used in the retrieval process, thus the entire process could be implemented end-to-end.
SDIM \cite{Cao2022Sampling} further improves ETA by proposing a simple hash sampling-based approach that directly gathers behavior items that share the same hash signature with the target item. 
ADFM \cite{Li2022Adversarial} proposes an adversarial filtering mechanism, which compresses the retrieved behavior sequences and removes the potential noisy behavior.

\subsection{Discussion}
The design philosophy of memory-augmented methods is to memorize more information by introducing more user-/item-specific parameters (i.e., external memory). 
Though effective in modeling longer sequences, the model is usually complicated and difficult to be deployed in real-world systems \cite{Pi2020Search}.
By comparison, retrieval-based frameworks are more efficient and easy to scale for large recommender systems.
These advantages make them a more suitable solution for handling long user behavior sequences in industrial applications.
However, retrieving behaviors inevitably causes information loss as many behaviors are dropped in the retrieval process. 
Therefore, it still remains an open problem regarding how to make a trade-off and exploit the advantages from both lines of research.

\section{Multi-Type UBM}\label{sec:mb-ubm}
Building on top of conventional UBM, multi-type UBM aims to explicitly consider the different behavior types as they convey subtle differences in user interest modeling \cite{Zhou2018Micro,Jin2020Multi}.
For example, in the e-commerce scenario, the \textit{purchase} behavior usually stands for stronger interests than the \textit{click} behavior.
While the \textit{positive} and \textit{negative} ratings convey opposite semantics in product reviews.
Specifically, multi-type UBM frameworks estimate the type-specific interest $\text{P}(u, i, b)$ of user $u$ in item $i$ as
\begin{equation}\label{eq:mb-form}
\text{P}(u, i, b) = F_{\Theta}^{UBM}(u, i, \mathcal{H}_u^{MB}), \ \forall \ u \in \mathcal{U}, \ i \in \mathcal{I},
\end{equation}
where $b$ is a specific behavior type in the whole behavior type set $\mathcal{B}$. 
The multi-type behavior history $\mathcal{H}_u^{MB}$ extends $\mathcal{H}_u$ by explicitly considering the behavior type in each behavior record, 
\begin{equation}\label{eq:mb-rec}
h_{u,k}^{MB} = \{v_k, t_k, b_k, f_k\}, \ b_k \in \mathcal{B},
\end{equation}
where $b_k$ is particularly picked out from the behavior-related feature $f_{k}$ to deal with the behavior multiplicity.
The explicit consideration of behavior types not only enables us to interpret user interests from a new perspective but also poses new challenges.

\subsection{Behavior Type Definition}
The first issue of multi-type UBM is the definition of behavior types, which is still an open question as the recommendation scenarios can vary drastically from each other. 
Researchers usually define different behavior types through expert analysis, which can be roughly divided into three categories.
\begin{itemize}[leftmargin=*]
    \item \textbf{Macro behaviors} refer to interactive behaviors with explicit motivations, which are defined by the logical design of recommendation scenarios.
    For example, the \textit{click}, \textit{add-to-cart}, \textit{add-to-favorite}, and \textit{buy} behaviors together implement the basic functions of e-commerce services \cite{Xia2021Knowledge,Xia2020Multiplex}.
    Users conduct a macro behavior to fulfill a distinctive purpose, which is well recorded (including the behavior type) and can be directly used for subsequent multi-type UBM methods.
    
    \item \textbf{Micro behaviors} are extracted from macro behaviors based on expert knowledge \cite{Zhou2018Micro,Meng2020Incorporating}, which provides a fine-grained understanding of user behaviors.
    For example, the \textit{click} behavior can be divided into different micro behaviors according to the click source, which helps to explain users' interests in specific items \cite{Zhou2018Micro}.
    However, definitions of micro behaviors are non-trivial and require sophisticated knowledge about both the particular recommendation scenarios and the targeted applications.
    
    \item \textbf{Behaviors from different domains or scenarios} can also be incorporated to provide complementary information for user interest mining \cite{Gu2021Self,Ma2019Pi}.
    For example, the ZEUS model combines the \textit{query} behavior in a search scenario with the \textit{click} behavior in a recommendation scenario to mine users' implicit feedback \cite{Gu2021Self}.
    Besides, the same \textit{watch} behavior in two domains is regarded as different behaviors to highlight the domain distinctions \cite{Ma2019Pi}.
    Depending on the applications, expert guidance is needed to select complementary behaviors from related domains or scenarios.
\end{itemize}

\subsection{Multi-Behavior Fusion}
After the behavior types are determined, the next challenge of multi-type UBM is how to further model the complicated cross-type behavior dependencies on top of conventional UBM models.
Existing researches fall into two groups according to how and when the cross-type behavior relations are fused with the intra-type behavior relations, i.e., the early-fusion models and the late-fusion models.

\begin{itemize}[leftmargin=*]
    \item \textbf{Late-fusion multi-type UBM} explores the intra-type and cross-type behavior relations under a two-step model architecture \cite{Gao2019Neural,Gu2020Deep,Gu2021Self,Chen2021Graph}.
    For example, the NMTR model first separately predicts the user-item interactions concerning different behavior types, then assembles them in a cascading manner to account for the cross-type behavior relations \cite{Gao2019Neural}.
    DMT and ZEUS propose to learn cross-type user representations based on the concatenation of intra-type sequential representations \cite{Gu2020Deep,Gu2021Self}.
    GHCF conducts multi-behavior predictions based on the behavior-wise fusion of intra-type collaborative representations \cite{Chen2021Graph}.
    Conventional UBM techniques (e.g., RNN, Transformer, and GCN) are readily usable for the intra-type learning process due to the homogeneity of behavior records. 
    Besides, the independent intra-type learning processes are parallelizable, which brings fast computational efficiency.
    However, the common limitation of late-fusion models is that they ignore the item-level cross-type behavior modeling, which may negatively impact the performance.

    \item \textbf{Early-fusion multi-type UBM} learns both intra-type and cross-type behavior relations jointly in a hybrid manner \cite{Zhou2018Micro,Yuan2022Multi,Guo2021DA,Wei2022Contrastive}.
    For example, in RIB \cite{Zhou2018Micro}, MB-STR \cite{Yuan2022Multi}, and $\pi$-Net \cite{Ma2019Pi} models, the multi-behavior sequential patterns are learned from the multi-type hybrid behavior sequences.
    While in MBGCN \cite{Jin2020Multi}, MB-GMN \cite{Xia2021Graph}, and MATN \cite{Xia2020Multiplex}, researchers attempt to explore the multiplex user-item interactive semantics with (heterogeneous) graph learning techniques.
    Building on top of conventional UBM, various modifications are proposed to incorporate behavior-aware information to deal with the item-level behavior differences. 
    On one hand, the performances are improved by the enumerated exploration of intra-type and cross-type behavior dependencies at the item level.
    On the other hand, the computation complexity also increases with the sophisticated model architecture.
\end{itemize}

\subsection{Multi-Behavior Prediction}
Another unique problem of multi-type UBM is that sometimes multiple types of behaviors are required to predict in the same model \cite{Gao2019Neural,Gu2020Deep,Gu2021Self,Guo2021DA}.
Joint prediction of different types of behaviors is challenging because the label distributions of different behaviors are not aligned in the same space or even mutually exclusive.
A common practice is to construct separated prediction modules for different behavior types, which is widely applicable regardless of the model paradigms \cite{Guo2019Buying,Guo2021DA,Chen2021Graph}.
However, the separated prediction head neglects the potential task relations, which leads to sub-optimal performances.
To avoid the negative transfer across different behavior prediction tasks, the MMoE \cite{Ma2018Modeling} and PLE \cite{Tang2020Progressive} methods are employed to promote task relevance and suppress task conflicts \cite{Gu2020Deep,Gu2021Self,Yuan2022Multi}.
When rigorous behavior dependencies are given by domain experts, the cascaded prediction structure can be applied \cite{Gao2019Neural}.

\subsection{Discussion}
The above three issues are closely related in multi-type UBM.
For example, the ZEUS model utilizes behavior types from different scenarios, which causes different label distributions and demands for sophisticated multi-behavior prediction modules \cite{Gu2021Self}.
DIPN simultaneously models macro and micro behaviors that express user interests at different levels, causing the early-fusion design incompetent to fuse them well \cite{Guo2019Buying}.
Therefore, multi-type UBM solutions should make comprehensive trade-offs between the above three key designs, tailoring for the specific application scenarios.

\section{UBM with Side Information}\label{sec:side-ubm}
Despite the great successes, most of the above-mentioned UBM methods overlook or underuse the rich side information associated with each behavior record.
To fill the gap, UBM with side information aims to design specialized components to exploit the rich knowledge contained in $f_k$,
\begin{equation}\label{eq:side-form}
\text{P}(u, i) = F_{\Theta}^{UBM}(u, i, M(\mathcal{H}_u)), \ \forall \ u \in \mathcal{U}, \ i \in \mathcal{I},
\end{equation}
where $M(\cdot)$ stands for the new side information fusing component.
When dealing with a behavior record $h_{u,k} = \{v_k, t_k, f_{k}\}$, $M(\cdot)$ fuses the side information ($t_k$ and $f_k$) with item ID ($v_k$) to obtain a fine-grained representation. 
Apparently, the design of $M(\cdot)$ relies heavily on the characteristics of adopted side information.

\subsection{Side Information Sources}
Various sources of side information can be utilized in UBM, which can be mainly classified into three categories, i.e., time information, item attributes, and multi-modal information.
In conventional UBM methods, such as SASRec \cite{Kang2018Self} and BERT4REC \cite{Sun2019BERT4Rec}, the time information is used to sort the behavior records, which only influences the position encoding before sequence modeling.
However, TiSASRec \cite{li2020time} finds that the time intervals between different item pairs convey crucial knowledge, thus proposing a novel time interval aware mechanism for the attentive weight calculation.
With a similar motive, TISSA \cite{lei2019tissa} firstly proposes a time interval-based GRU to obtain session-level behavior representations, then splits them into slices with multi-scale time windows to better capture temporal dependencies.
The item attributes also influence user behaviors significantly, providing necessary context information.
FDSA \cite{zhang2019feature} proposes to combine the item ID with attributes like category, brand, and description text for sequential recommendations.
To track the changes of attributes over time, Trans2D \cite{singer2022sequential} performs feature transformation over item ID and attributes to learn complex item-attribute patterns.
Different from the above methods, the use of multi-modal side information is more complicated.
p-RNN \cite{hidasi2016parallel} first extracts image features from video thumbnails and text features from product descriptions.
Then it adopts existing CV and NLP approaches (GoogLeNet and bag-of-words) to learn multi-modal feature representations separately.
Inspired by the achievements of multi-modal pre-training, SEMI \cite{lei2021semi} obtains video and text representations through the direct use of pre-trained CV and NLP SOTAs.

\subsection{Side Information Utilization}
Once obtaining the side information representations, the effective use of them is another key issue.
Early works usually merge different representation vectors with the simple addition or concatenation operations, then feed the mixed item vector to subsequent learning modules.
For example, p-RNN \cite{hidasi2016parallel} proposes to fuse the item ID information and side information at the input level with a concatenated feature or at the output level with a weighted summation output. 
RNN units are applied for sequence processing. 
SC-CNN \cite{zhang2022deepvt} regards side information as different views, concatenates them as a 3D cube, then uses a semi-causal CNN to simultaneously capture the relations of different views.
Trans2D \cite{singer2022sequential} also transforms items with attributes as a 3D cube, but applies a modified Transformer with 2D self-attention to handle the 3D data.
CARCA \cite{rashed2022context} further extends the self-attention network with a two-branch multi-head self-attention-based framework to capture the dynamic user preferences hidden in users' context and attribute-aware profiles.
The left branch is used to extract the dependencies among the rich historical behaviors and profiles, and the right branch is used to capture the influence of the behavior sequence on the target item.
However, fusing side information with simple fusion operators may negatively impact the original item ID representations.
As a result, NOVA-BERT \cite{liu2021noninvasive} proposes to leverage side information as an auxiliary for the self-attention module to learn better attention distribution, instead of being fused into item representations.
DIF-SR \cite{xie2022decoupled} argues that integrating side information before attention calculation will limit
the learning of attention matrices thus decouples various side information with separate attention calculations to further improve NOVA-BERT.
Inspired by the success of Self-Supervised Learning (SSL) in CV and NLP fields, researchers also adopt SSL techniques for side information modeling.
To mine the item-attribute relations, S3Rec \cite{zhou2020s3} proposes two attribute-related SSL objectives in pre-training, i.e., associated attribute prediction and masked attribute prediction.
To further consider item attributes at the interest level, MISS \cite{Guo2022MISS} proposes a CNN-based extractor to capture interest representations, dependencies, and correlations.

\subsection{Discussion}
It can be observed that the heterogeneous sources of side information play a decisive role in the subsequent extraction and fusion processes.
The simple time and attribute information can be seamlessly integrated into behavior modeling, while the complex image and text information needs to be processed with existing techniques in an ad-hoc manner.
Thus, there is a lot of room to explore how to efficiently and effectively integrate the side information into the interest representation space.

\section{Industrial Practices}\label{sec:industrial}

\begin{table*}[t]
\centering
\begin{tabular}{l|l|l|l}
\hline
Model                 & Application Scenario                & Baseline       & Gains \& Costs \\
\hline
DIN                   & Online advertising                  & Embedding\&MLP & +10.0\% CTR, +3.8\% RPM \\
\hline
\multirow{2}{*}{DIEN} & \multirow{2}{*}{Online advertising} & Embedding\&MLP & +20.7\% CTR, +17.1\% eCPM, -3.0\% PPC \\
                      &                                     & DIN            & +11.8\% CTR, +10.4\% eCPM, -1.0\% PPC\\
\hline
GRU4Rec+              & Online video                        & Strategy       & +5\% Watch time, +5\% Video play, +4\% Click \\
\hline
\multirow{1}{*}{BST}  & \multirow{1}{*}{E-commerce}
                                                            & DIN            & +3.02\% CTR, +4ms RT \\
\hline
\specialrule{0em}{0pt}{2pt}
\multicolumn{4}{c}{\small{(a) Conventional UBM}} \\
\specialrule{0em}{0pt}{3pt}
\hline
UIC                   & Online advertising                  & DIEN           & +7.5\% CTR, +6\% RPM \\
\hline
\multirow{2}{*}{SIM}  & \multirow{2}{*}{Online advertising} & UIC            & +7.1\% CTR, +4.4\% RPM, +53 times MSL \\
                      &                                     & DIEN           & +2.1 $d_{category}$ \\
\hline
UBR                   & App store                           & w/o UBR        & +6.6\% eCPM, +11.1\% CTR \\
\hline
ETA                   & E-commerce                          & SIM            & +1.8\% CTR, +3.1\% GMV, -2ms IT \\
\hline
SDIM                  & Online search                       & w/o Long sequence & +2.98\% CTR, +2.69\% VBR, +1ms IT \\
\hline
ADFM                  & Online advertising                  & SIM            & +4.7\% CTR, +3.1\% RPM, -70.8\% Storage \\
\hline
\specialrule{0em}{0pt}{2pt}
\multicolumn{4}{c}{\small{(b) Long-sequence UBM}} \\
\specialrule{0em}{0pt}{3pt}
\hline
\multirow{1}{*}{DMT}  & \multirow{1}{*}{E-commerce}
                                                            & DIEN           & +4.5\% CTR, +4.6\% CVR, +6.0\% GMV \\
\hline
\multirow{1}{*}{ZEUS} & \multirow{1}{*}{E-commerce}         
                                                            & DMT            & +6.0\% CTR, +9.7\% CVR, 11.7\% GMV \\
\hline
DIPN                  & Coupon allocation                   & Strategy       & +41.1\% Usage Rate, +39.8\% GMV \\
\hline
\specialrule{0em}{0pt}{2pt}
\multicolumn{4}{c}{\small{(c) Multi-type UBM}} \\
\specialrule{0em}{0pt}{3pt}
\hline
NOVA-BERT             & App store                           & BERT           & +$0.192\times 10^9$ FLOPs, +7.1 Mb Model Size\\
\hline
SEMI                  & E-commerce                          & BST            & +9.32\% NBV, +10.45\% DT, +12.10\% CWR \\
\hline
TiSSA                 & E-commerce                          & w/o TiSSA      & +1.56\% CTR, +2.09\% CVR, +3.66\% GMV \\
\hline
\specialrule{0em}{0pt}{2pt}
\multicolumn{4}{c}{\small{(d) UBM with side information}} \\
\specialrule{0em}{0pt}{3pt}
\hline
\multicolumn{4}{l}{CTR: Click-Through Rate; RPM: Revenue Per Mille; eCPM: effective Cost Per Mille; PPC: Pay Per Click; RT: Response} \\ 
\multicolumn{4}{l}{Time; MSL: Max Sequence Length; $d_{category}$: Days till Last Same Category Behavior; IT: Inference Time; VBR: Visited} \\
\multicolumn{4}{l}{Buy Rate; CVR: ConVersion Rate; GMV: Gross Merchandise Value; NBV: Number of Browsing Videos; DT: Dwell Time;} \\
\multicolumn{4}{l}{CWR: Complete-Watch Ratio.} \\
\hline
\specialrule{0em}{0pt}{2pt}
\multicolumn{4}{c}{\small{(e) Explanation of evaluation metrics}} \\
\specialrule{0em}{0pt}{3pt}
\end{tabular}
\vspace{-5mm}
\caption{Industrial online deployments of UBM solutions claimed in published papers.}
\label{tab:apply}
\vspace{-5mm}
\end{table*}

The research of recommender systems is highly application-oriented, which is challenged by various practical problems.
Therefore, pure research-oriented design is not sufficient enough to reflect the potential values of UBM research.
In this section, we discuss the industrial practices of representative UBM methods.
To ensure data authority and accuracy, online performance statics are collected from published papers only.
The results are presented in Table \ref{tab:apply}, from which we have the following observations.

\begin{itemize}[leftmargin=*]
    \item Besides the frequently mentioned online advertising and e-commerce, UBM methods are also applicable in other scenarios, such as APP store and coupon allocation, which reflects the potential research and application value of UBM.
    
    
    \item The consideration of long-range behaviors, multi-type behaviors, and side information all achieve improvements over conventional UBM baselines (e.g., UIC v.s. DIEN, DMT v.s. DIEN, SEMI v.s. BST), which demonstrates the value of the two key directions.
    However, if we consider the number of successfully deployed models, long-sequence UBM takes the lead.
    
    \item Computational efficiency is another key concern in UBM industrial practices.
    Despite that BST outperforms WDL and DIN in terms of CTR, the Transformer unit also causes a higher response time (RT), which limits its application to high throughput scenarios.
    When it comes to long behavior sequences, hashing (ETA), sampling (SDIM), and denoising (ADFM) methods demonstrate their superiority in balancing between performances and costs.
    
    \item Despite the popularity of GNN techniques in academic research (e.g., SURGE, MBGCN, MB-GMN, etc.), none of them have been deployed online.
    The process of graph data requires a large amount of computational resources, which causes a heavy burden for the online environment.
\end{itemize}

\section{Summary and Future Prospects}\label{sec:future}
This survey summarizes the recent advances in user behavior modeling.
Rich user preferences can be discovered from the behavior logs.
With the help of RNN, CNN, and Attention-based techniques, conventional UBM methods are capable of finding the implicit feedback from the behavior sequences, thus largely improving the recommendation performances.
Further improvements are obtained by taking into account the long-term behavior histories, the multi-type behaviors, and the side information accompanied with user behaviors.
Despite the appreciable achievements, however, UBM research still faces some challenges from both academic and industrial perspectives.

\begin{itemize}[leftmargin=*]
    \item \textbf{Deeper information fusion.} Despite the existing explorations on long behavior sequences, multi-type behaviors, and side information, the combination of them has seldom been considered.
    For example, modeling long multi-type behavior sequences might lead to new improvements, but also demands novel type-aware retrieval or sampling techniques.
    Therefore, there still exists much exploration space for the wider and deeper fusion of various useful information.
    
    \item \textbf{More efficient learning method.} As application-oriented research, there exists a trade-off between effectiveness and efficiency for online serving.
    Along with the growing behavior length, behavior types, and kinds of side information, the computational complexity and storage burden of UBM solutions also increase significantly.
    With the purpose of maintaining compelling performances, it is always important to seek UBM solutions with a light computation burden.
    
    \item \textbf{More interpretable user representations.} Despite the achieved improvements, the learned user interest representations are not well interpretable, which restricts the use case of UBM to other applications, such as user profiling and causal analysis.
    Thus interpretable UBM is another promising future research direction.
    
    \item \textbf{More advanced techniques.} Advanced deep learning techniques like pre-training and big models have been academically explored for side information learning.
    However, there is still a big gap before these models can be deployed in industrial environments.
    
\end{itemize}

\bibliographystyle{named}
\bibliography{ijcai23}

\end{document}